\documentclass[12pt,preprint]{aastex}

\shorttitle{Five Recycled Pulsars}
\shortauthors{Jacoby et al.}

\begin{document}
 
\title{Discovery of Five Recycled Pulsars in a High Galactic Latitude Survey}

\author{B. A. Jacoby\altaffilmark{1,2}, M. Bailes\altaffilmark{3},
S. M. Ord\altaffilmark{3,4}, H. S. Knight\altaffilmark{3,5}, and
A. W. Hotan\altaffilmark{3,5}}

\altaffiltext{1}{Department of Astronomy, California Institute of
Technology, MS 105-24, Pasadena, CA 91125.}

\altaffiltext{2}{present address: Naval Research Laboratory, Code 7213,
4555 Overlook Avenue, SW, Washington, DC 20375; bryan.jacoby@nrl.navy.mil.}

\altaffiltext{3}{Centre for Astrophysics and Supercomputing, Swinburne
University of Technology, P.O. Box 218, Hawthorn, VIC 31122,
Australia; mbailes@astro.swin.edu.au, hknight@astro.swin.edu.au, ahotan@astro.swin.edu.au.}

\altaffiltext{4}{present address: School of Physics, University of Sydney, A28,
NSW 2006, Australia; ord@physics.usyd.edu.au.}

\altaffiltext{5}{Australia Telescope National Facility, CSIRO,
P.O. Box 76, Epping, NSW 1710, Australia.}

\begin{abstract}
We present five recycled pulsars discovered during a 21-cm survey of
approximately 4,150 deg$^2$ between 15$\arcdeg$ and 30$\arcdeg$ from
the galactic plane using the Parkes radio telescope.  One new pulsar,
PSR~J1528$-$3146, has a 61\,ms spin period and a massive white
dwarf companion. Like many recycled pulsars with heavy companions, the
orbital eccentricity is relatively high ($\sim$0.0002), consistent
with evolutionary models that predict less time for circularization.
The four remaining pulsars have short spin periods ($3\,{\rm ms} < P <
6\,{\rm ms}$); three of these have probable white dwarf binary
companions and one (PSR J2010$-$1323) is isolated. PSR J1600$-$3053 is
relatively bright for its dispersion measure of 52.3\,pc\,cm$^{-3}$ and
promises good timing precision thanks to an intrinsically narrow
feature in its pulse profile, resolvable through coherent
dedispersion.  In this survey, the recycled pulsar discovery rate was
one per four days of telescope time or one per 600 deg$^2$ of
sky.  The variability of these sources implies that there are more
millisecond pulsars that might be found by repeating this survey.
\end{abstract}


\section{Introduction}\label{sec:intro}

Most pulsars are thought to descend from massive stars in the disk of
the galaxy, and as a result, the bulk of pulsar search efforts have
historically been concentrated near the galactic plane where the
majority of pulsars, with their fairly short observable lifetimes,
reside \citep[e.g.][]{sstd86,clj+92,jlm+92,mlc+01}.  Early surveys
along the plane met with limited success when searching for recycled
pulsars.  The value of finding new recycled pulsars comes from their
contribution to our understanding of the binary evolution processes
\citep{bv91}, tests of general relativity \citep{vbb+01}, and their
use in a millisecond pulsar (MSP) timing array for the detection of
low frequency gravitational waves \citep{fb90}.

The pioneering survey of \cite{wol91a} led to the discovery of two
fascinating systems in a small area at high galactic latitude.  This
success spawned several surveys at high galactic latitudes, which
yielded a large number of recycled pulsars
\citep{bhl+94,fcwa95,lnl+95,nll+95,cnt96}.  Simulations of the
low-luminosity galactic recycled pulsar population point to a more
isotropic distribution than is expected for the more luminous
long-period pulsars that are less subject to the deleterious effects
of the interstellar medium \citep{jb91}.  This difference is because
recycled pulsars have had a longer time to migrate away from their
birthplace in the galactic disk and their shorter periods strongly
limit the distance to which they can be detected in the electron-rich
galactic plane.  These facts, combined with the relative insensitivity
to dispersion and lower sky temperatures afforded by high-frequency
observations, suggested that a 21\,cm survey for pulsars away from the
galactic plane would be extremely productive \citep{tbms98}.  The
success of the Swinburne Intermediate Latitude Pulsar Survey
demonstrated the validity of this approach, discovering eight recycled
pulsars in $\sim 2950$\,deg$^2$ at galactic latitudes between
5$^\circ$ and 15$^\circ$ \citep{eb01, eb01b, ebvb01}. Recent
re-processing of the Parkes multibeam survey which concentrated on the
galactic plane ($\left| b \right| < 5^\circ$) has also yielded
millisecond and recycled pulsars in large numbers \citep{fsk+04}.

In this paper we describe five recycled pulsars discovered in this
extension of the Swinburne Intermediate Latitude survey.  In section
\ref{sec:survey} we describe the survey parameters before discussing
the five pulsars individually in section 3, giving their timing
solutions (where possible), pulse profiles and derived parameters. In
section 4 we discuss our results.

\section{A High Latitude Pulsar Survey}\label{sec:survey}

This survey was carried out using the 13-beam multibeam receiver on
the Parkes 64\,m radio telescope from January 2001 to December 2002.
This survey covered $\sim4,150$\,deg$^2$ in the region $-100 ^\circ <
l < 50 ^\circ$, $15 ^\circ < \left| b \right| < 30 ^\circ$.
Relatively short 265~s integrations gave a sensitivity which is
well-matched to the expected scale height and luminosity distribution
of the pulsar population \citep{cc97} and allowed us to complete the
7232 survey pointings in about four weeks of observing.  The signals
from each beam were processed and digitized by a $2 \times 96\times
3$\,MHz filterbank operating at a center sky frequency of 1374\,MHz
and one-bit sampled every 125\,$\mu$s, providing good sensitivity to
fast pulsars with low to moderate dispersion measures.  This observing
methodology is identical to that employed for the Swinburne
Intermediate Latitude Pulsar Survey and differs from that of the
Parkes Multibeam Pulsar Survey \citep{mlc+01} only in sampling period
(125\,$\mu$s vs. 250\,$\mu$s) and integration time (265\,s
vs. 2100\,s).

The resulting 2.4~TB of data were searched for pulsar-like signals
using standard techniques with the 64 Compaq Alpha workstations at the
Swinburne Centre for Astrophysics and Supercomputing, resulting in the
discovery of 26 new pulsars.  Full details of this survey will be
described in a future paper (Jacoby et al. in preparation).  Of these
26 new pulsars, seven are recycled.  One of them, PSR~J1909$-$3744, is
an exceptionally interesting millisecond pulsar and has been reported
elsewhere \citep{jbvk+03,jhb+05}, and another, PSR~J1738+0333, will be
described in a latter paper (Jacoby et al. in preparation). The other
five are described here.

\section{Discovery and Timing of Five Recycled Pulsars}\label{sec:sevenpsrs}

\begin{figure}
\plotone{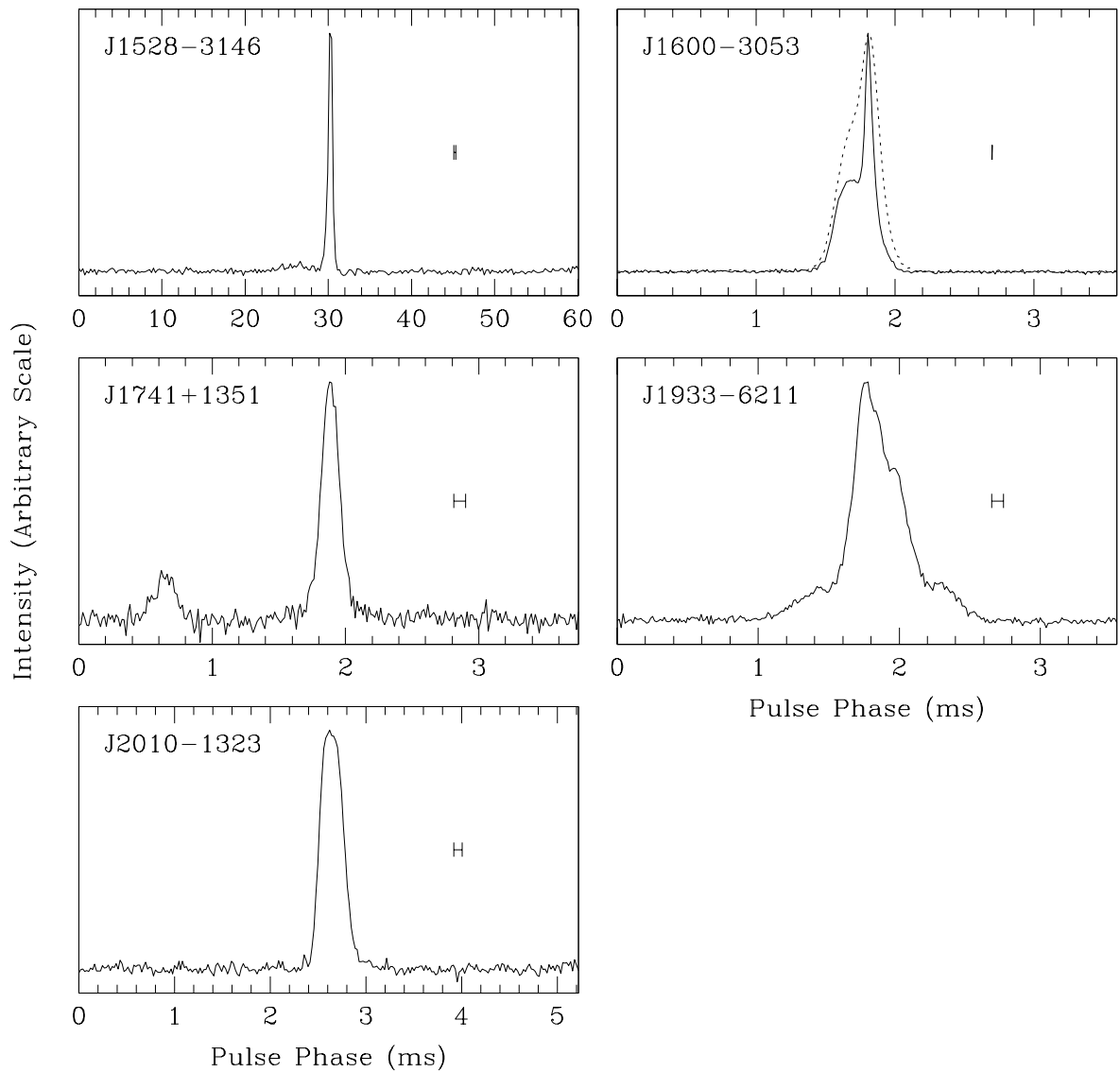}
\caption{Average pulse profiles at 1.4~GHz.  PSR~J1600-3053 profile
measured with CPSR2 (solid) and the $512 \times 0.5$\,MHz filterbank
(dotted), and J1741+1351 profile measured by the $96 \times 3$\,MHz
filterbank.  The $512 \times 0.5$\,MHz filterbank was used for all
others.  Horizontal bars represent the time resolution of the
observing system arising from the differential dispersion within a
filterbank channel and the sampling interval, except for J1600$-$3053
where horizontal bar indicates 2\,$\mu$s time resolution of the coherently
dedispersed pulse profile.}
\label{fig:profiles}
\end{figure}

These five objects belong to the class of recycled pulsars with
relatively weak magnetic fields and small spindown rates.  Four are
MSPs with spin periods well under 10\,ms; the fifth has a longer spin
period, but is still clearly recycled due to its large characteristic
age ($\tau_c$).  One of the MSPs is isolated; the other pulsars are in
binary systems with probable white dwarf companions.  Average pulse
profiles of the five pulsars are shown in Figure~\ref{fig:profiles}.

We have begun a systematic timing program at Parkes for these and
other pulsars discovered in this survey, primarily using the $2 \times
512 \times 0.5$\,MHz filterbank at 1390\,MHz with occasional
observations using the $2 \times 256 \times 0.125$\,MHz filterbank at
660\,MHz to determine the dispersion measure (DM).

We followed standard pulsar timing procedures: folded pulse profiles
from individual observations were cross-correlated with a high
signal-to-noise template profile to determine an average pulse time of
arrival (TOA) corrected to UTC(NIST).  The standard pulsar timing
package {\sc tempo}\footnote{http://pulsar.princeton.edu/tempo}, along
with the Jet Propulsion Laboratory's DE405 ephemeris, was used for all
timing analysis.  TOA uncertainties for each pulsar were multiplied by
a factor between 1.12 and 1.65 to achieve reduced $\chi^2 \simeq 1$.
One of the new pulsars, PSR~J1933$-$6211, has a very
small orbital eccentricity ($e$), giving rise to a strong covariance
between the time of periastron ($T_0$) and longitude of periastron
($\omega$) in this system.  For this pulsar, we have used the ELL1
binary model which replaces $\omega$, $T_0$, and $e$ with the time of
ascending node ($T_{\rm asc}$) and the Laplace-Lagrange parameters $e
\sin \omega$ and $e \cos \omega$ \citep{lcw+01}.  We have used the DD model 
\citep{dd85, dd86} for PSR~J1528$-$3146 and PSR~J1600$-$3053.   

For PSR~J1741+1351, orbital parameters were obtained by fitting a
model to observed spin periods obtained from multiple observations
with the 96-channel survey filterbank.  The position reported here is
the center of the survey beam in which the pulsar was discovered;
timing analysis of this pulsar is ongoing and will be reported in a
future paper (Freire et al. in preparation).

\begin{figure}
\plotone{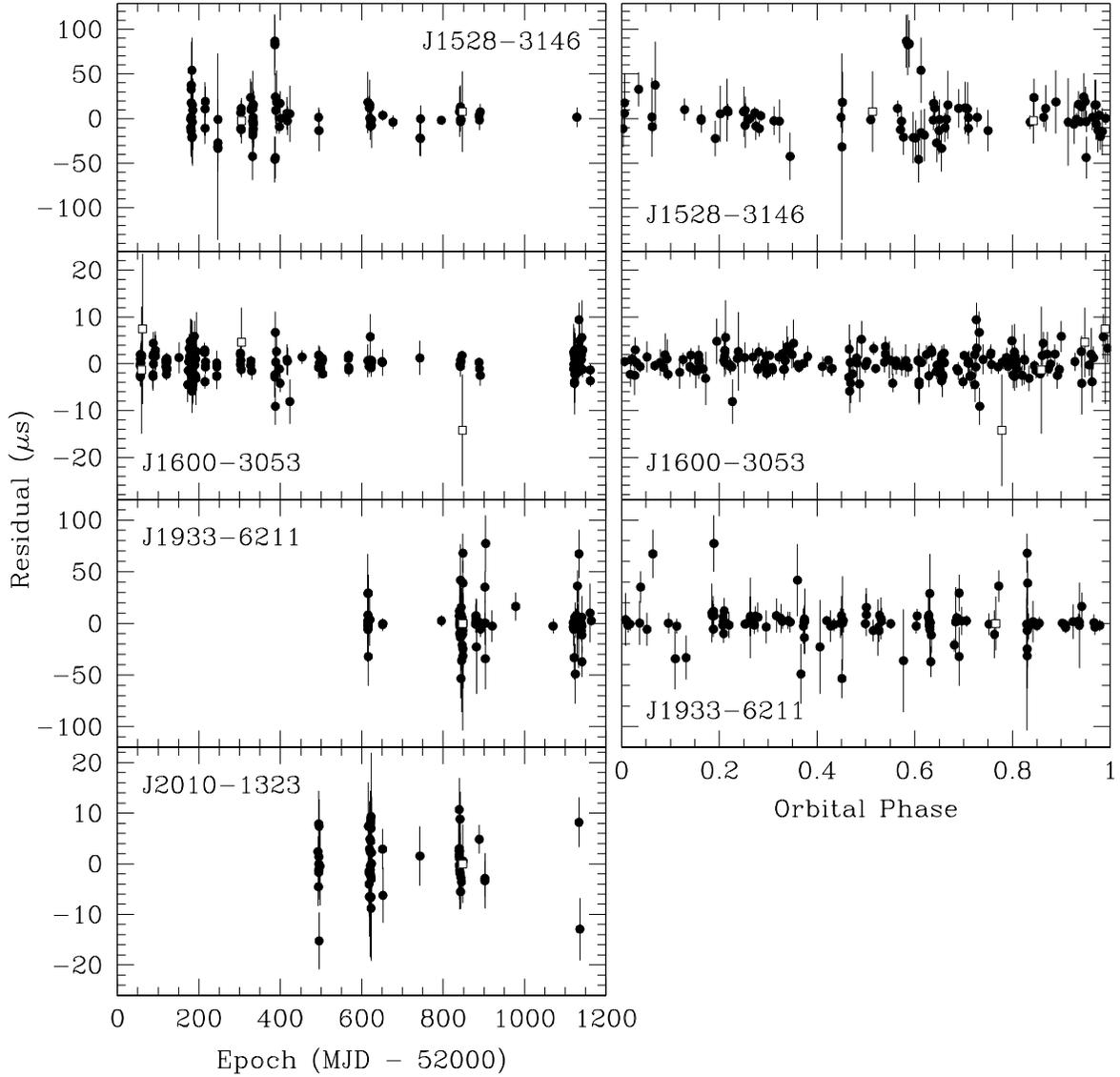}
\caption{Timing residuals plotted versus observation epoch (left column) and orbital phase (right column) for pulsars with phase-connected timing solutions.  Filled circles represent observations at 1390\,MHz, open squares represent 600\,MHz observations.}
\label{fig:residuals}
\end{figure}

Astrometric, spin, binary, and derived parameters for all pulsars are
given in Tables \ref{tab:par1} and \ref{tab:par2}.  Timing residuals
for the four pulsars with timing solutions are shown in Figure
\ref{fig:residuals}.

\subsection{PSR~J1528$-$3146: A Recycled Pulsar with a Massive Companion}

PSR J1528$-$3146 has the longest spin period ($P = 61$\,ms) of the five
new pulsars reported here.  The minimum companion mass (obtained from
the mass function by assuming an edge-on orbit and a pulsar mass of
1.35\,M$_{\odot}$) is 0.94\,M$_{\odot}$; the system's circular orbit
suggests that the companion must be a CO or ONeMg white dwarf.  Only a
handful of such intermediate mass binary pulsar (IMBP) systems are
known \citep{eb01b,clm+01}.  Of all low-eccentricity binary pulsars,
this object has the second highest minimum companion mass and fifth
highest projected orbital velocity.  The orbital parameters of
PSR~J1528$-$3146 are broadly similar to those of PSR~J1157$-$5112, the
low-eccentricity binary pulsar with the most massive white dwarf
companion \citep{eb01}. The orbital eccentricity ($\sim$2x10$^{-4}$)
is high compared to recycled pulsars with similar orbital periods. The
large mass of the companion suggests a shorter duration spin-up phase,
leaving less time to circularize the post-supernova orbit and transfer
mass to the pulsar. At the dispersion measure of 18\,pc\,cm$^{-3}$,
the implied distance is only about 1\,kpc.  We have detected a
potential optical counterpart at the pulsar timing position with
$R \sim 24.2$.  As with other optically-detected IMBP companions, this
is brighter than expected given the characteristic age of the pulsar,
suggesting that the spin-down age overestimates the cooling age of the
white dwarf and hence the time since the end of mass transfer in the
system \citep{jcvk+06}.

\subsection{PSR~J1600$-$3053: A High Precision Timing LMBP}

Our timing results obtained with the Parkes filterbank for this
3.6\,ms pulsar have a weighted RMS residual of 1.55\,$\mu$s over
three years.  Using the Caltech-Parkes-Swinburne Recorder II
\citep[CPSR2; see ][]{jac05}, a wide-bandwidth coherent dedispersion
backend at Parkes, \citet{ojh+06} have achieved an RMS residual of
only 650\,ns for this pulsar; measurement of the Shapiro delay constrains the orbital inclination to be between 59\arcdeg and 70\arcdeg~(95\% confidence).  This
pulsar will be an important part of pulsar timing array experiments
aimed at detecting low-frequency gravitational waves emitted by
coalescing supermassive black hole binaries \citep{jb03}.

\subsection{PSR~J1741+1351: An LMBP with Strong Scintillation}

This 3.7\,ms LMBP scintillates very strongly at 1.4\,GHz, frequently
making it undetectable in reasonable ($\sim$15 min) integration times
at Parkes.  In fact, four attempts were required to confirm this
pulsar candidate highlighting why repeating this survey would no doubt
turn up other pulsars that are, on average, below the nominal survey
sensitivity limit.  This strong scintillation partly explains why this
pulsar was not discovered in previous Arecibo surveys.  We have not
yet obtained a phase-connected timing solution for this pulsar. The
minimum companion mass of 0.24 M$_\odot$, 16 day orbital period and
inconsistent flux make it uninteresting for tests of General
Relativity with current instrumentation.

\subsection{PSR~J1933$-$6211: An MSP with an Edge-on Orbit?}

This short period (3.4 ms) pulsar's companion has a minimum mass of
0.32\,$M_\odot$ --- somewhat higher than the typical LMBP.  The only
other disk pulsars with companions of at least 0.3\,$M_\odot$ and spin
periods shorter than 10\,ms are PSR~J2019+2425 \citep{ntf93}
($P = 3.9$\,ms) with a much longer orbital period ($P_b = 76$\,d
vs. 13\,d), and PSR~J1757$-$5322 \citep{eb01b} ($P = 8.9$\,ms) and
PSR~J1435$-$6100 \citep{clm+01} ($P = 9.3$\,ms) with substantially more
compact orbits ($P_b = 1.4$\,d and 0.45\,d respectively). This may
mean that this pulsar has a nearly edge-on orbit or represents a limit
to accretion spin-up given the mass of the companion.

\subsection{PSR~J2010$-$1323: Isolated MSP}

This 5.2\,ms pulsar is the only isolated MSP found in this survey.
Approximately one-fourth of recycled pulsars not associated with
globular clusters are isolated. The RMS timing residual is 4 $\mu$s
with a 256 MHz filterbank system, and there is no evidence of any
planetary system like that surrounding the planet pulsar PSR B1257+12.
With hour-long integrations and coherent dedispersion backend, we
anticipate that a weighted RMS resdiual of approximately $1\,\mu s$
will be possible for this pulsar with the Parkes telescope and perhaps
half that with the 100\,m Green Bank Telescope.

\section{Discussion}

This survey has once again demonstrated the efficacy of the Parkes
64\,m telescope's multibeam receiver for discovering recycled pulsars.
The recycled pulsar discovery rate of one per 600 deg$^2$ is
comparable to that of Edwards et al. (2001).  However, only one new
recycled pulsar was discovered more than 25$\arcdeg$ and less than
30$\arcdeg$ from the plane, compared to the five between 5$\arcdeg$
and 10$\arcdeg$ from the plane, which may indicate that the density of
MSPs detectable with this backend and telescope combination drops off
considerably when more than 25$\arcdeg$ from the galactic plane.

The ratio of the number of recycled pulsars still in binaries compared
to those that are now single puts a limit on the fraction of systems
where recycling goes out of control, resulting in a solitary pulsar
either because the pulsar gets too close to the companion and ablates
it with its strong wind, or tidally disrupts the companion completely.
In this survey 6 of the 7 recycled pulsars are in binary systems,
which in itself suggests that in the majority of cases recycling is
not fatal to the donor star. Thanks to the Parkes multibeam surveys
the disk population of millisecond pulsars is now becoming large
enough for us to search for trends that might explain why some
millisecond pulsars are single and others are not, and which
parameters are correlated. The recycled pulsars found in this survey
have reinforced the previously observed trend of pulsars with
high-mass companions having longer spin periods and higher
eccentricities than those with low-mass companions.

Large-scale uniform surveys allow us to model the underlying
populations' properties, including the $z$-scale height of MSPs. There
are 11 field MSPs (defined for our purposes as having $P < 20$\,ms)
with $|z| > 0.5$\,kpc in the ATNF Pulsar
Catalogue\footnote{http://www.atnf.csiro.au/research/pulsar/psrcat/}
\citep{mht+05}, with $DM$-based distances provided by the NE2001 model
\citep{cl02} for pulsars without direct distance estimates.  Of these
pulsars, 4 were discovered by the Parkes 70\,cm survey \citep{mld+96,
lml+98}, and none by the Swinburne surveys (though we would have
expected these surveys to have similar sensitivities to high-$|z|$
pulsars).  The Arecibo surveys that are responsible for most of the
rest are difficult to model.  To investigate the implications of the
large-area Parkes surveys we simulated a population of 100,000 MSPs
with a wide range in luminosity and a scale height of 500\,pc.  Our
simulation predicted that half of all discoveries should have $|z| >
0.5$\,kpc for the Parkes 70cm and the Swinburne surveys. In reality,
only about 13\% of the MSPs detected have $|z| >$ 0.5\,kpc, which
either suggests that 0.5 kpc is an overestimate of their scale height,
or our modeling of their luminosities is incorrect.  We note that
somewhat different results are obtained using the older \citet{tc93}
Galactic electron model to estimate distances.  Based on this model,
there are 17 MSPs with $|z| > 0.5$\,kpc, 5 of which were discovered in
the Parkes 70\,cm survey and 5 of which were discovered by the
Swinburne surveys.  This suggests that values of the $z$ scale height
from previous population studies based on the older distance model,
such as the $0.65^{+0.16}_{-0.12}$\,kpc Gaussian scale height of
\citet{cc97}, will be revised downward by the use of the new Galactic
electron model.

\begin{figure}
\plotone{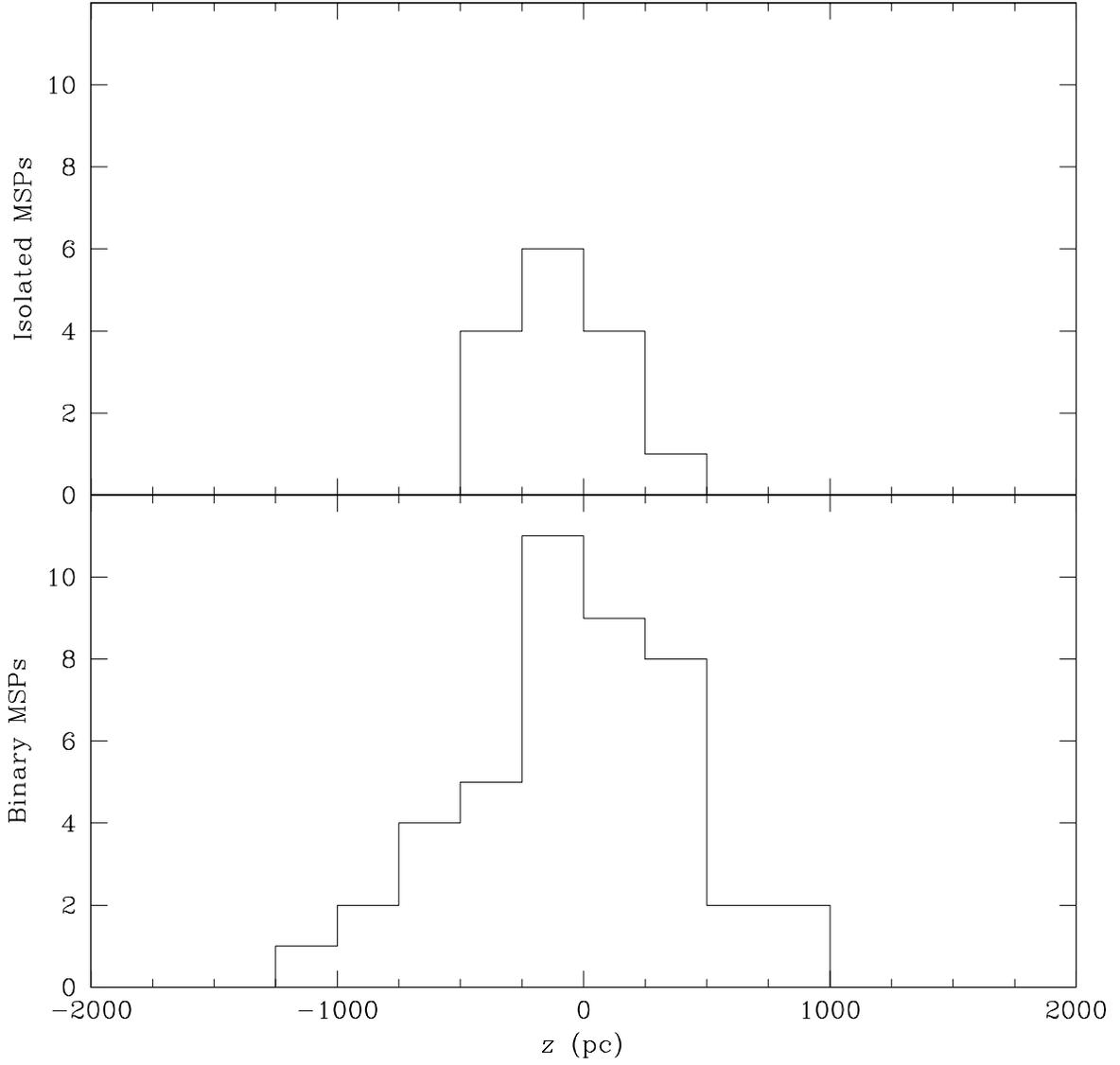}
\caption{Histograms of height above the Galactic plane for isolated (upper panel) and binary (lower panel) field MSPs.}
\label{fig:zhist}
\end{figure}

Again restricting ourselves to disk pulsars with $P < 20$\,ms, we find 15
isolated MSPs and 44 in binary systems.  The distribution of height
above the Galactic plane is shown for these two populations in Figure
\ref{fig:zhist}.  The RMS deviation from $z = 0$ is $240 \pm 40$\,pc
for isolated MSPs and $450 \pm 50$\,pc for binary MSPs (if we consider
only pulsars with $P < 10$\,ms the isolated sample is unchanged, while
the binary sample drops to 37 objects with RMS in $z$ of $470 \pm
60$\,pc).  These results are similar to those obtained by
\citet{lkn+06} using a sample of 9 isolated and 20 binary pulsars with
$P < 10$\,ms.

The median distance to the isolated and binary MSPs in our sample is
960\,pc and 1240\,pc respectively, regardless of the period cutoff
used.  The difference between these two populations is less than that
noted by \citet{lkn+06}, who found more than a factor of two
difference in their sample.  This result may suggest that the
luminosity distributions of isolated and binary MSPs are not as
different as previously thought, with implications for the birth
velocity distributions required by the different observed scale
heights.  A flux monitoring campaign on a large population of MSPs
including those discovered here should enable direct investigation of
the tantalising scale height---luminosity question for isolated and
binary MSPs. If the isolated MSPs really do have lower luminosities we
would expect them to have had a different accretion history than those
still in binary systems.  If it is the tight binary MSPs that evolve
into isolated MSPs through either ablation or tidal destruction of
their companions, then we might expect different underlying velocity
distributions and therefore different scale heights for the isolated
and binary populations.

\acknowledgments

We thank R.~Edwards for invaluable help with pulsar search software.
  The Parkes telescope is part of the Australia Telescope which is
  funded by the Commonwealth of Australia for operation as a National
  Facility managed by CSIRO.  BAJ thanks NSF and NASA for supporting
  this research.  BAJ holds a National Research Council Research
  Associateship Award at the Naval Research Laboratory (NRL).  Basic
  research in radio astronomy at NRL is supported by the Office of
  Naval Research.


\begin{deluxetable}{lrr}
\tablecaption{Pulsar Parameters for J1528$-$3146 and J1600$-$3053\label{tab:par1}}
\tablecolumns{3}
\tabletypesize{\scriptsize}
\tablewidth{0pt}
\tablehead{
    \colhead{Parameter\tablenotemark{a}} 
      & \colhead{J1528$-$3146}
      & \colhead{J1600$-$3053}
}
\startdata
Right ascension, $\alpha_{\rm J2000}$\dotfill
   & $15^{\rm h}28^{\rm m}34\fs9542(2)$
   & $16^{\rm h}00^{\rm m}51\fs90392(2)$
\\
Declination, $\delta_{\rm J2000}$\dotfill
   & $-31\arcdeg46\arcmin06\farcs836(8)$
   & $-30\arcdeg53\arcmin49\farcs325(2)$
\\
Proper motion in $\alpha$, $\mu_{\alpha}$ (mas yr$^{-1}$)\dotfill
   & \nodata
   & -0.91(51)
\\
Proper motion in $\delta$, $\mu_{\delta}$ (mas yr$^{-1}$)\dotfill
   & \nodata
   & -4.0(15)
\\Pulse period, $P$ (ms)\dotfill
   & 60.82223035146(1)
   & 3.59792845222642(6)
\\
Reference epoch (MJD)\dotfill
    & 52500.0
    & 52500.0
\\
Period derivative, $\dot{P}$ (10$^{-20}$)\dotfill
   & 24.9(1)
   & 0.9479(4)
\\
Dispersion measure, DM (pc cm$^{-3}$)\dotfill
   & 18.163(6)
   & 52.333(1)
\\
Binary model\dotfill
   & DD
   & DD
\\
Binary period, $P_b$ (d)\dotfill
   & 3.180345754(3)
   & 14.348457554(4)
\\
Projected semimajor axis, $a \sin i$ (lt-s)\dotfill
   & 11.452324(5)
   & 8.8016571(4)
\\
Orbital eccentricity, $e$\dotfill
   & 0.000213(1)
   & 0.00017371(8)
\\
Longitude of periastron, $\omega$ (deg)\dotfill
   & 296.83635$\pm$0.2
   & 181.768043$\pm$0.03
\\
Time of periastron, $T_0$\dotfill
   & 52502.4013744$\pm$0.002
   & 52506.3711244$\pm$0.001
\\
Weighted RMS timing residual ($\mu$s)\dotfill &
11.1 & 1.55 
\\
\cutinhead{Derived Parameters}
Minimum companion mass $m_{c~{\rm min}}$ (M$_{\odot}$)\dotfill
   & 0.94
   & 0.20
\\
Galactic longitude, $l$ (deg)\dotfill
    & 337.94 
    & 344.09
\\
Galactic latitude, $b$ (deg)\dotfill
    & 20.22
    & 16.45
\\
DM-derived distance, $d$ (kpc)\tablenotemark{b}\dotfill
    & 0.80
    & 1.53
\\
Distance from Galactic plane, $|z|$ (kpc)\dotfill
    & 0.28
    & 0.43
\\
Transverse velocity, $v_{\perp}$ (km s$^{-1}$)\tablenotemark{c}\dotfill
    & \nodata
    & $30^{+16}_{-15}$
\\
Surface magnetic field, $B_{\rm surf}$ (10$^8$\,G)\dotfill
    & 39.3
    & 1.8\tablenotemark{d}
\\
Characteristic age, $\tau_c$ (Gyr)\dotfill
   & 3.9  
   & 6.2\tablenotemark{d}
\\
Pulse FWHM, $w_{50}$ (ms)\dotfill &
 0.59 & 0.079 
\\
Pulse width at 10\% peak, $w_{10}$ (ms)\dotfill &
 1.29 & 0.41 
\\
Discovery S/N \dotfill &
 28.0 & 16.7 
\\
Flux Density, $S_{1400}$ (mJy)\tablenotemark{e} \dotfill &
 1.1 & 3.2
\\
\enddata
\tablenotetext{a}{Figures in parenthesis are uncertainties in the last digit quoted.  Uncertainties are calculated from twice the formal error produced by {\sc tempo}.}
\tablenotetext{b}{From the model of \citet{cl02}.}
\tablenotetext{c}{Stated uncertainty in transverse velocity is based only on uncertainty in proper motion; the distance is taken as exact.}
\tablenotetext{d}{Corrected for secular acceleration based on measured proper motion and estimated distance \citep{shk70}.}
\tablenotetext{e}{Flux density estimated from observed S/N and nominal system parameters, except for PSR~J1600$-$3053 from Ord et al. (2004).}
\end{deluxetable}
\nocite{ovs+04}

\begin{deluxetable}{lrrr}
\tablecaption{Pulsar Parameters for J1741+1351, J1933$-$6211, and J2010$-$1323 \label{tab:par2}}
\tablecolumns{4}
\tabletypesize{\scriptsize}
\tablewidth{0pt}
\tablehead{
    \colhead{Parameter\tablenotemark{a}} 
      & \colhead{J1741+1351}
      & \colhead{J1933$-$6211}
      & \colhead{J2010$-$1323}
}
\startdata
Right ascension, $\alpha_{\rm J2000}$\dotfill
   & $17^{\rm h}41^{\rm m}37^{\rm s}\pm1^{\rm m}$
   & $19^{\rm h}33^{\rm m}32\fs4272(3)$
   & $20^{\rm h}10^{\rm m}45\fs9196(2)$
\\
Declination, $\delta_{\rm J2000}$\dotfill
   & $+13\arcdeg54\arcmin41\arcsec\pm14\arcmin$
   & $-62\arcdeg11\arcmin46\farcs881(4)$
   & $-13\arcdeg23\arcmin56\farcs027(6)$
\\Pulse period, $P$ (ms)\dotfill
   & 3.7471544(6)
   & 3.3543431438847(1)
   & 5.223271015190(1)
\\
Reference epoch (MJD)\dotfill
   & \nodata 
   & 53000.0
   & 52500.0
\\
Period derivative, $\dot{P}$ (10$^{-20}$)\dotfill
   & \nodata
   & 0.37(1)
   & 0.482(7)
\\
Dispersion measure, DM (pc cm$^{-3}$)\dotfill
   & 24.0(3)
   & 11.499(7)
   & 22.160(2)
\\
Binary model\dotfill
   & \nodata
   & ELL1
   & \nodata
\\
Binary period, $P_b$ (d)\dotfill
   & 16.335(2)
   & 12.81940650(4)
   & \nodata
\\
Projected semimajor axis, $a \sin i$ (lt-s)\dotfill
   & 11.03(6)
   & 12.281575(3)
   & \nodata
\\
$e \sin \omega$ ($\times 10^{-6}$)\dotfill
   & \nodata
   & 1.1(4)
   & \nodata
\\
$e \cos \omega$ ($\times 10^{-6}$)\dotfill
   & \nodata
   & -0.55(50)
   & \nodata
\\Time of ascending node, $T_{\rm asc}$ (MJD)\dotfill 
   & 52846.22(1)
   & 53000.4951005(5)
   & \nodata
\\
Weighted RMS timing residual ($\mu$s)\dotfill &
\nodata & 6.06 & 4.25
\\
\cutinhead{Derived Parameters}
Orbital eccentricity, $e$\dotfill
   & \nodata
   & 0.0000013(4)
   & \nodata
\\
Longitude of periastron, $\omega$ (deg)\dotfill
   & \nodata
   & 115.93036$\pm$22
   & \nodata
\\
Time of periastron, $T_0$ (MJD)\dotfill
   & \nodata
   & 53004.6233183$\pm$0.8
   & \nodata
\\
Minimum companion mass $m_{c~{\rm min}}$ (M$_{\odot}$)\dotfill
   & 0.24
   & 0.32
   & \nodata
\\
Galactic longitude, $l$ (deg)\dotfill
    & 37.94
    & 334.43
    & 29.45
\\
Galactic latitude, $b$ (deg)\dotfill
    & 21.64
    & -28.63
    & -23.54
\\
DM-derived distance, $d$ (kpc)\tablenotemark{b}\dotfill
    &  0.91
    &  0.52
    &  1.02
\\
Distance from Galactic plane, $|z|$ (kpc)\dotfill
    & 0.34
    & 0.25
    & 0.41
\\
Surface magnetic field, $B_{\rm surf}$ (10$^8$\,G)\dotfill
    & \nodata
    & 1.2
    & 1.6
\\
Characteristic age, $\tau_c$ (Gyr)\dotfill
   & \nodata
   & 15
   & 17
\\
Pulse FWHM, $w_{50}$ (ms)\dotfill &
 0.16 & 0.36 & 0.28
\\
Pulse width at 10\% peak, $w_{10}$ (ms)\dotfill &
 0.32 & 1.03 & 0.44 
\\
Discovery S/N \dotfill &
 10.7 & 22.9 & 12.4
\\
Flux Density, $S_{1400}$ (mJy)\tablenotemark{c} \dotfill &
 0.93 & 2.3 & 1.6
\\\enddata
\tablenotetext{a}{Figures in parenthesis are uncertainties in the last digit quoted.  Uncertainties are calculated from twice the formal error produced by {\sc tempo}.}
\tablenotetext{b}{From the model of \citet{cl02}.}
\tablenotetext{c}{Flux density estimated from observed S/N and nominal system parameters.}
\end{deluxetable}

\end{document}